\documentclass[10pt]{article}
\usepackage{graphicx}
\usepackage[bottom=1in,left=1in,right=1in,top=1in]{geometry}

\date{September 28, 2005}
\title{Transverse Momentum Distributions and String Percolation Study in p+p, d+Au and Au+Au at $\sqrt{s_{NN}}$ = 200 GeV}
\author{Terence J. Tarnowsky*, Brijesh K. Srivastava, Rolf P. Scharenberg\\
(for the STAR Collaboration)\\\\
Purdue University\\
Department of Physics\\
525 Northwestern Ave\\	
West Lafayette, IN 47907, USA\\
Phone: 1-(765)-494-9776\\
Fax: 1-(765)-494-0706\\
\ $^{*}$Corresponding Author: tjt@physics.purdue.edu}
\begin{document}

\maketitle
\begin{abstract}
Multiparticle production at high energies is described in terms of color strings stretched between the projectile and target. As string density increases, overlap among the strings leads to cluster formation. At some critical density a macroscopic cluster appears, spanning the entire system. This marks the percolation phase transition. Data from p+p, d+Au and Au+Au collisions at 200 GeV has been analyzed using the STAR detector to obtain the percolation density parameter, $\eta$. For 200 GeV Au+Au collisions, the value of $\eta$ is found to lie above the critical percolation threshold, while for 200 GeV d+Au collisions it is below the critical value. This supports the idea of string percolation, which at high enough string density is a possible mechanism to explore the hadronic phase transition to a quark-gluon plasma. \end{abstract}

\twocolumn
\section{Introduction}
It is postulated that in the collision of two nuclei at high-energy, color strings are formed between projectile and target partons. These color strings decay into additional strings via  \begin{math} 
	q-\overline{q}
\end{math}
production, and ultimately hadronize to produce the observed hadron yields \cite{Intro}.
\\
In the collision process, partons from different nucleons begin to overlap and form clusters in transverse space. The color strings are of radius r$_{0}$ = 0.20-0.25 fm \cite{Intro}. \\The fusion of strings to form clusters is an evolution of the Dual Parton Model \cite{Intro2}, which utilizes independent strings as particle emitters, to the Parton String Model (PSM), which implements interactions (fusion) between strings \cite{Intro3}. At some point, a cluster will form which spans the entire system. This is referred to as the maximal cluster and marks the onset of the percolation threshold. An overview of percolation theory can be found in the following reference \cite{Intro4}. The quantity $\eta$, the percolation density parameter, can be used to describe overall cluster density. It can be expressed as 
\begin{equation}\eta = \frac{N\pi r_{0}^{2}}{S}\end{equation} 
with N the number of strings, S the total nuclear overlap area, and $\pi r_{0}^{2}$ the transverse disc area. At some critical value of $\eta = \eta_{c}$, the percolation threshold is reached. $\eta_{c}$ is referred to as the critical percolation density parameter. In two dimensions, for a uniform string density, and in the continuum limit, $\eta_{c}$ = 1.175 \cite{Intro5}.
\\
To calculate the percolation parameter, $\eta$, a parameterization of pp events at 200 GeV is used to compute the $p_{T}$ distribution 
\begin{equation}\frac{dN}{dp_{T}^{2}} = \frac{a}{(p_{0}+p_{T})^{n}}\end{equation}
where a, $p_{0}$, and n are parameters fit to the data. This parameterization can be used for nucleus-nucleus collisions if one takes into account the percolation of strings by \cite{Intro}
\begin{equation}p_{0} \longrightarrow p_{0}\left( \frac{\left<\frac{nS_{1}}{S_{n}}\right>_{Au-Au}}{\left<\frac{nS_{1}}{S_{n}}\right>_{pp}}\right)^{\frac{1}{4}}\end{equation}
\\
In pp collisions at 200 GeV, the quantity $\left<\frac{nS_{1}}{S_{n}}\right>_{pp} = 1.0 \pm 0.1$, due to low string overlap probability in pp collisions. Once the $p_{T}$ distribution for nucleus-nucleus collisions is determined, the multiplicity damping factor can be defined in the thermodynamic limit as \cite{Intro6}
\begin{equation}\label{eq4}F(\eta) = \sqrt{\frac{1-e^{-\eta}}{\eta}}\end{equation}
which accounts for the overlapping of discs, with $1-e^{-\eta}$ corresponding to the fractional area covered by discs. 
\section{Data Analysis}
The data utilized in this analysis was acquired by the STAR experiment at RHIC. Only data from within $\pm$ 1 unit of pseudorapidity with greater than 10 fit points in the main STAR detector, the Time Projection Chamber, was used. The events studied were minimum bias events, with standard STAR centrality cuts as defined by the trigger detectors and overall event multiplicities, which increase with increasing collision centrality. The longitudinal (z) vertex was constrained to within $\pm$ 30 centimeters of the primary event vertex. The distance of closest approach (dca) of primary tracks to the event vertex was less than 3.0 cm. The transverse momentum of charged particles was less than 1.2 GeV.
\\
The collision systems studied includes: Au+Au at 200, 62.4, and 19.6 GeV and pp, d+Au, and Cu+Cu at 200 GeV. 
\section{Results and Discussion}

The percolation density parameter, $\eta$, has been determined for several collision systems and energies. These results have been compared to the predicted value of the critical percolation density, $\eta_{c}$. If $\eta_{c}$ is exceeded it is expected that the percolation threshold has been reached, indicating the formation of a maximal cluster that spans the system under study. 
\\
\begin{figure}
\centering
\includegraphics[width=3in]{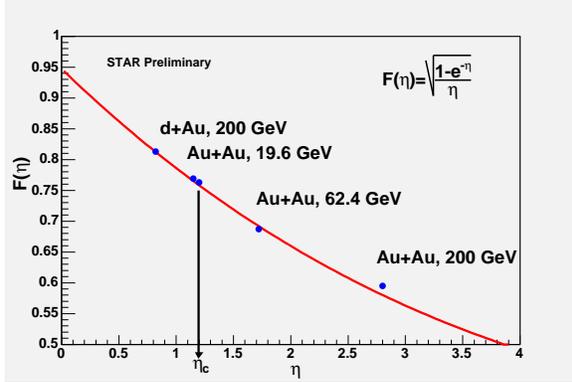}
\caption{\footnotesize{Multiplicity suppression factor, F($\eta$) versus the percolation density parameter, $\eta$. The line is the function F($\eta$) and is drawn to guide the eye, not as a fit to the points. The estimated critical percolation density for 2-D overlapping discs in the continuum limit, $\eta_{c}$, is shown. The $\eta$ values are calculated for the most central collisions. The values for 200 GeV d+Au and 19.6 GeV Au+Au lie below $\eta_{c}$, while those for 62.4 and 200 GeV Au+Au lie above.}}
\label{Fig1}
\end{figure}
Figure 1 is a plot of the quantity F($\eta$) (Eq.$\ref{eq4}$) versus the percolation density parameter, ($\eta$), for central collisions. The red line is a representation of F($\eta$) and is drawn to guide the eye. The critical percolation threshold is indicated as that for overlapping strings with uniform density \cite{Intro5}. There is a general increase in $\eta$ with increasing system size and energy. The predicted value of $\eta_{c}$ lies just above that of 19.6 GeV Au+Au collisions and well above that of 200 GeV d+Au. Both 200 and 62.4 GeV Au+Au lie well above the predicted percolation threshold. It is expected that the overlap probability for string clusters increases as the system size and energy increase. Therefore, a larger percolation density should be obtained at 200 GeV Au+Au compared to lower energies and other, lighter systems.
\\
One can also consider the percolation density as a function of centrality in Au+Au collisions. The centrality expressed in terms of the number of participating nucleons ($N_{part}$) as found from Monte Carlo Glauber calculations \cite{Glauber}. More central collisions correspond to greater values of $N_{z}$. Figure 2 shows $\eta$ as a function of the number of participant nucleons in Au+Au collisions at 200 and 62.4 GeV. It is shown here that for all centralities, except for the most peripheral bin, all 200 GeV Au+Au collisions lie above the critical percolation threshold. The value of $\eta$ increases with increasing collision centrality, an expected indication of additional string overlap in more central collisions. For 62.4 GeV Au+Au, almost all centralities, except for the three most peripheral bins, lie above the critical percolation density. Another interesting feature is that the value of $\eta$ appears to achieve a maximum value of approximately 1.8 in central 62.4 GeV Au+Au collisions, whereas in central 200 GeV Au+Au collisions $\eta$ continues to increase. This could potentially indicate a saturation of the cluster overlap area, and hence the percolation parameter in 62.4 GeV Au+Au collisions.   

\begin{figure}
\centering
\includegraphics[width=3in]{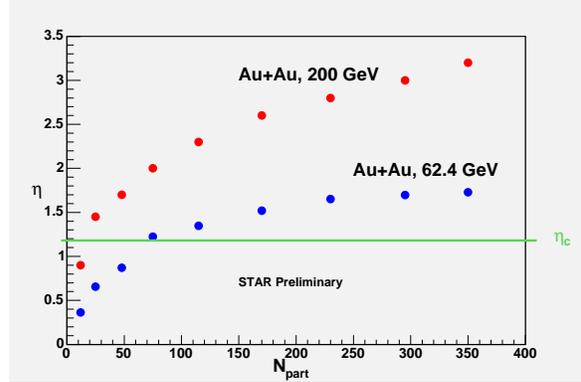}
\caption{\footnotesize{The percolation density parameter, $\eta$, as a function of collision centrality ($N_{part}$) in 62.4 and 200 GeV Au+Au collisions. Larger values of $N_{part}$ correspond to more central collisions. For almost all collision centralities, 200 GeV Au+Au exceeds the critical percolation threshold, $\eta_{c}$. In 62.4 GeV Au+Au, all except the three most peripheral bins exceed $\eta_{c}$.}}
\label{Fig2}
\end{figure}

Finally, Fig. 3 presents the fractional area covered by clusters as a function of $\eta$. The data is shown for all centralities of 62.4 and 200 GeV Au+Au collisions, as well as new data with 200 GeV Cu+Cu results. The 200 GeV Cu+Cu result covers a region between that of mid-central to central 62.4 GeV Au+Au and mid-peripheral 200 GeV Au+Au. Because $\eta$ is dependent on energy and atomic number, this behavior is naively expected. Unlike the Au+Au results, the Cu+Cu data is not corrected for efficiency and detector acceptance. It has been predicted by Satz \cite{Intro5} that the study of Cu+Cu collisions at RHIC energies provides one of the best possibilities to study the onset of deconfinement due to the fact that for heavier nuclei (such as Pb+Pb) at top RHIC energy, all centralities are above the percolation threshold. As shown in Fig. 3, 200 GeV Cu+Cu and 62.4 GeV Au+Au appear to probe the critical percolation threshold.

\section{Summary}
In summary, we have presented the preliminary results for the percolation density parameter, $\eta$, at RHIC for several collision systems as a function of collision centrality. Percolation theory describes the hadronic to partonic phase transition as occurring at a critical percolation density, $\eta_{c}$. These results show that for light systems (d+Au) or heavy systems at low energies (Au+Au at 19.6 GeV), the calculated value for $\eta$ lies below the predicted critical value. For minbias Au+Au collisions at 62.4 and 200 GeV, $\eta$ lies well above the predicted percolation threshold. As a function of centrality in Au+Au collisions at 62.4 and 200 GeV, almost all centralities except the most peripheral for top RHIC energy are above the percolation threshold, while for the lower 62.4 GeV events, several peripheral bins lie just at, or below, $\eta_{c}$. We have also shown preliminary results for Cu+Cu collisions at top RHIC energy compared to 62.4 and 200 GeV Au+Au results. The Cu+Cu results span the regime between central 62.4 and peripheral 200 GeV Au+Au. With increasing $\eta$, the fractional area of the system covered by clusters increases. This behavior is expected due to the overall increase in the number of strings as a function of collision energy and atomic number of the colliding nuclei. These results indicate that 200 GeV Cu+Cu and 62.4 GeV Au+Au may provide the ability to study the onset of the critical percolation threshold as a function of collision centrality. The onset of this collective behavior in traditional percolation theory is indicative of a phase transition, which may provide clues to whether a QGP is formed in heavy-ion collisions at RHIC.\\

\begin{figure}
\centering
\includegraphics[width=3in]{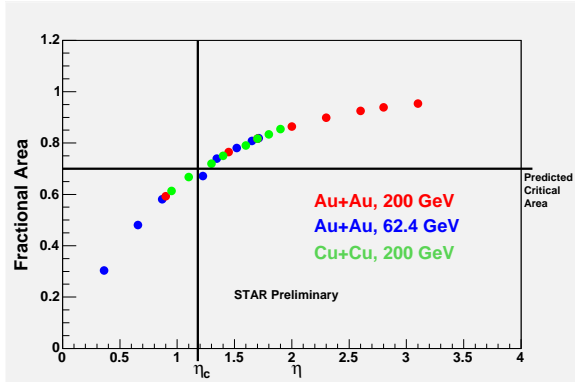}
\caption{\footnotesize{The fractional area covered by clusters as a function of the percolation density parameter, $\eta$. Results for 200 and 62.4 GeV Au+Au, along with 200 GeV Cu+Cu are shown. The new, preliminary Cu+Cu results lay in the regime that spans the range of $\eta$ covered by peripheral 200 GeV Au+Au and central 62.4 GeV Au+Au.}}
\label{Fig3}
\end{figure}

\begin{center}
\textbf{Acknowledgements}\\
\end{center}
We thank the RHIC Operations Group and RCF at BNL, and the NERSC Center at LBNL for their support. This work was supported in part by the HENP Divisions of the Office of Science of the U.S. DOE; the U.S. NSF; the BMBF of Germany; IN2P3, RA, RPL, and
EMN of France; EPSRC of the United Kingdom; FAPESP of Brazil; the Russian Ministry of Science and Technology; the Ministry of Education and the NNSFC of China; IRP and GA of the Czech Republic, FOM of the Netherlands, DAE, DST, and CSIR of the Government of India; Swiss NSF; the Polish State Committee for Scientific Research; STAA of Slovakia, and the Korea Sci. \& Eng. Foundation. We would also like to thank Nestor Armesto and Carlos Pajares for their fruitful discussions.

\bibliographystyle{unsrt}
\bibliography{test}
\end{document}